# UBR+: Improving Performance of TCP over ATM-UBR service[1]

Rohit Goyal, Raj Jain, Shiv Kalyanaraman, Sonia Fahmy
Department of Computer and Information Science, 395 Dreese Lab
2015 Neil Avenue
The Ohio State University
Columbus, OH 43210-1277
E-mail: {*goyal,jain,shivkuma,fahmy*}@cis.ohio-state.edu
Phone: (614)-292-3989. Fax: (614)-292-2911

Seong-Cheol Kim
Samsung Electronics Co. Ltd.
Chung-Ang Newspaper Bldg.
8-2, Karak-Dong, Songpa-Ku
Seoul, Korea 138-160
Email: kimsc@metro.telecom.samsung.co.kr

## Abstract

ATM-UBR switches respond to congestion by dropping cells when their buffers become full. TCP connections running over UBR experience low throughput and high unfairness. For 100% TCP throughput each switch needs buffers equal to the sum of the window sizes of all the TCP connections. Intelligent drop policies can improve the performance of TCP over UBR with limited buffers. The UBR+ service proposes enhancements to UBR for intelligent drop. Early Packet Discard improves throughput but does not attempt to improve fairness. Selective packet drop based on per-connection buffer occupancy improves fairness. The Fair Buffer Allocation scheme further improves both throughput and fairness.

## 1 Introduction

The Unspecified Bit Rate (UBR) service provided by ATM networks has no explicit congestion control mechanisms [8]. However, it is expected that many TCP implementations will use the UBR service category. TCP employs a window based end-to-end congestion control mechanism to recover from segment loss and also avoid congestion collapse. Several studies have been done to analyze the performance of TCP over the UBR service [1, 4, 11]. TCP sources running over ATM switches with limited buffers experience low throughput and high unfairness [2, 3, 7, 10].

Studies have shown that intelligent drop policies at switches can improve throughput of transport connections. Early Packet Discard (EPD) [1] proposed by Romanov and Floyd has been shown to improve TCP throughput but not fairness [7]. A policy for selective cell drop based on per-VC accounting can be used to improve fairness. Enhancements that perform intelligent cell drop policies at the switches need to be developed for UBR to improve transport layer throughput and fairness.

Heinanen and Kilkki [6] have designed a drop policy called Fair Buffer Allocation (FBA) that attempts to improve fairness among connections. The FBA scheme selectively drops complete packets from a connection based on the connection's buffer occupancy. The scheme uses a FIFO buffer at the switch, and performs some per-VC accounting to keep track of each VC's buffer occupancy. FBA tries to allocate a fair share of bandwidth to competing sources by managing the amount of buffer space used by each connection.

In this paper, we analyze several enhancements to the ATM UBR service category. This enhanced service category is called UBR+ because it maintains the simplicity of UBR and performs congestion control without explicit feedback control mechanisms. UBR+ improves throughput and fairness by intelligent cell drop policies. We describe the performance of TCP over UBR and its various enhancements.

---
[1] Submitted to ICC'97, 8-12 June 1997, Montreal.
Available from http://www.cis.ohio-state.edu/~jain/papers/icc97.ps

We first discuss the congestion control mechanisms in the TCP protocol and explain why these mechanisms can result in low throughput during congestion. We then describe our simulation setup used for all our experiments and define our performance metrics. We present the performance of TCP over vanilla UBR and explain why TCP over vanilla UBR results in poor performance. We then describe the Early Packet Discard scheme and present simulation results of TCP over UBR with EPD. Next, we present a simple selective drop policy based on per-VC accounting. This is a simpler version of the Fair Buffer Allocation scheme as proposed by Heinanen and Kilkki. We present an analysis of the operation of these schemes and the effect of their parameters. We also provide guidelines for choosing the best FBA parameters.

## 2 TCP congestion control

TCP relies on a window based protocol for congestion control. TCP connections provide end-to-end flow control to limit the number of packets in the network. The flow control is enforced by two windows. The receiver's window (RCVWND) is enforced by the receiver as measure of its buffering capacity. The Congestion Window (CWND) is kept at the sender as a measure of the capacity of the network. The sender sends data one window at a time, and cannot send more than the minimum of RCVWND and CWND into the network.

The TCP congestion control scheme consists of the "Slow Start" and "Congestion Avoidance" phases. The variable SSTHRESH is maintained at the source to distinguish between the two phases. The source starts transmission in the slow start phase by sending one segment (typically 512 Bytes) of data, i.e., CWND = 1 TCP segment. When the source receives an acknowledgement for a new segment, the source increments CWND by 1. Since the time between the sending of a segment and the receipt of its ack is an indication of the Round Trip Time (RTT) of the connection, CWND is doubled every round trip time during the slow start phase. The slow start phase continues until CWND reaches SSTHRESH (typically set to 64K bytes) and then the congestion avoidance phase begins. During the congestion avoidance phase, the source increases its CWND by 1/CWND every time a segment is acknowledged. The slow start and the congestion avoidance phases correspond to an exponential increase and a linear increase of the congestion window every round trip time respectively.

If a TCP connection loses a packet, the destination responds by sending duplicate acks for each out-of-order packet received. The source maintains a retransmission timeout for the last unacknowledged packet. The timeout value is reset each time a new segment is acknowledged. Congestion is detected by the source by the triggering of the retransmission timeout. At this point, the source sets SSTHRESH to half of CWND. More precisely, SSTHRESH is set to $\max\{2, \min\{CWND/2, RCVWND\}\}$. CWND is set to one.

As a result, CWND < SSTHRESH and the source enters the slow start phase. The source then retransmits the lost segment and increases its CWND by one every time a new segment is acknowledged. The source proceeds to retransmit all the segments since the lost segment before transmitting any new segments. This corresponds to a go-back-N retransmission policy. Note that although the congestion window may increase beyond the advertised receiver window (RCVWND), the source window is limited by the minimum of the two . The typical changes in the source window plotted against time are shown in Figure 1.

Most TCP implementations use a 500 ms timer granularity for the retransmission timeout. The TCP source estimates the Round Trip Time (RTT) of the connection by measuring the time (number of ticks of the timer) between the sending of a segment and the receipt of the ack for the segment. The retransmission timer is calculated as a function of the estimates of the average and mean-deviation of the RTT [12]. Because of coarse grained TCP timers, when there is loss due to congestion, significant time may be lost waiting for the retransmission timeout to trigger. The source does not send any new segments when duplicate acks are being received. When the retransmission timeout triggers, the connection enters the slow start phase. As a result, the link may remain idle for a long time and experience low utilization. Moreover, the sender attempts to retransmit all the segments since the lost segment. Many of these may be discarded at the destination if the latter had cached the out-of-order segments.

**Coarse granularity TCP timers and retransmission of segments by the go-back-N policy are the main reasons that TCP sources can experience low throughput and high file transfer delays during congestion.**

TCP Reno includes the Fast Retransmit and Fast Recovery algorithms that improve TCP performance when a single segment is lost. However, in high bandwidth links, network congestion can result in several dropped segments. In this case, fast retransmit and recovery are not able to recover from the loss and slow start is triggered. In our experiments, typical losses are due to congestion and result in multiple segments being dropped. Therefore, we study TCP without fast retransmit and recovery running on UBR.

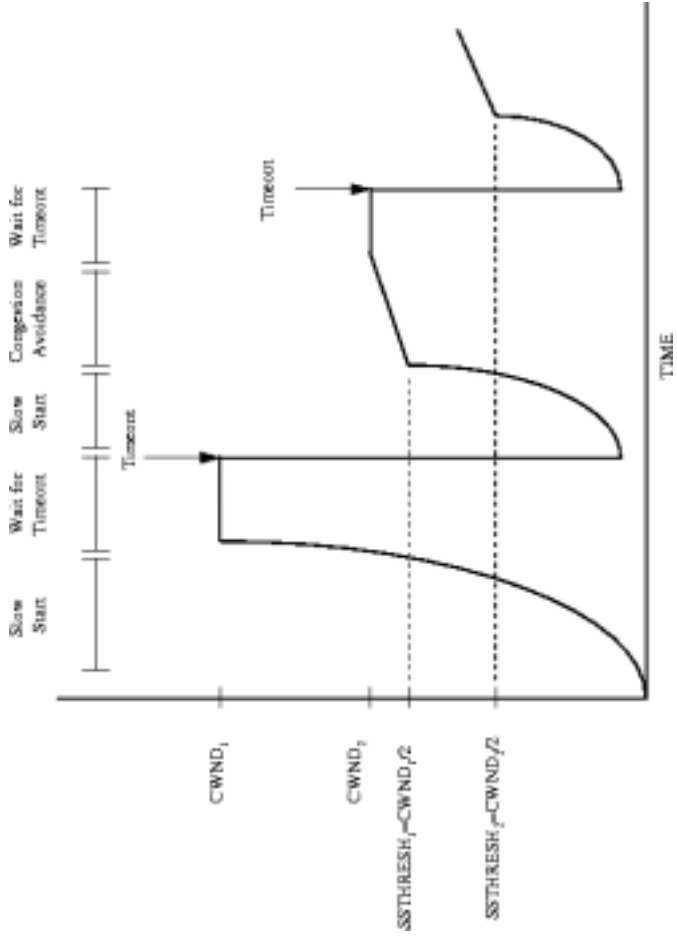

Figure 1: TCP CWND vs Time

## 3 The Simulation Experiment

### 3.1 Simulation Model

All simulations presented in this paper are performed on the N source configuration shown in Figure 2. The configuration consists of N identical TCP sources that send data whenever allowed by the window. The switches implement UBR service with optional drop policies described in this paper. The following simulation parameters are used [11]:

- The configuration consists of N identical TCP sources as shown in Figure 2.

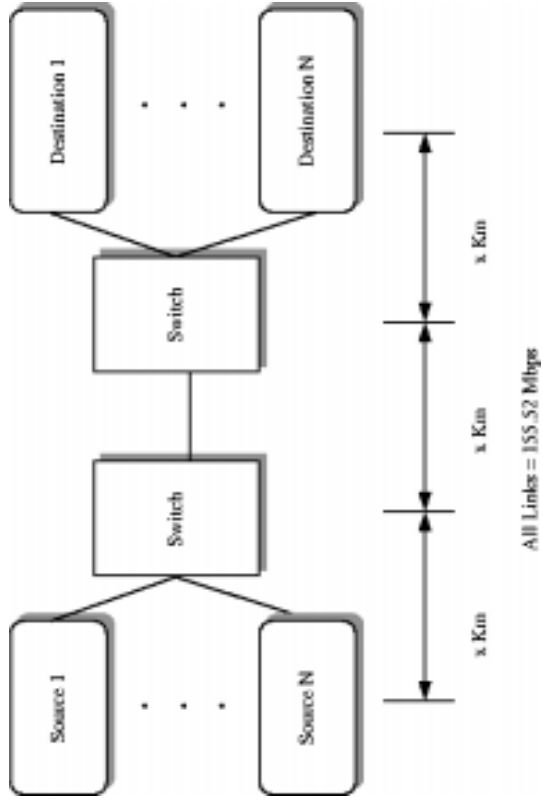

Figure 2: The N-source TCP configuration

- All sources are infinite TCP sources. The TCP layer always sends a segment as long as it is permitted by the TCP window.

- All link delays are 5 microseconds for LANs and 5 milliseconds for WANs.

- All link bandwidths are 155.52 Mbps.

- Peak Cell Rate is 155.52 Mbps.

- The traffic is unidirectional. Only the sources send data. The destinations send only acknowledgments.

- TCP Fast Retransmit and Recovery are disabled. This isolates the slow-start and congestion avoidance behavior of TCP. Moreover, Fast Retransmit and Recovery are unable to handle multiple packet loss, which is seen in our simulations.

- The TCP segment size is set to 512 bytes. This is the standard value used by current TCP implementations. Larger segment sizes have been reported to produce higher TCP throughputs, but these have not been implemented in real TCP protocol stacks.

- TCP timer granularity is set to 100 ms. This affects the triggering of retransmission timeout due to packet loss. The values used in most TCP implementations is 500 ms, and some implementations use 100 ms. Several other studies have used smaller TCP timer granularity and have obtained higher throughput numbers. However, the timer granularity is an important factor in determining the amount of time lost during congestion. Small granularity results in less time being lost waiting for the retransmission timeout to trigger. This results in faster recovery and higher throughput. However, TCP implementations do not use timer granularities of less than 100 ms, and producing results with lower granularity artificially increases the throughput.

- TCP maximum receiver window size is 64K bytes for LANs. This is the default value used in TCP. For WANs, this value is not enough to fill up the pipe, and reach full throughput. In the WAN simulations we use the TCP window scaling option to scale the window to the bandwidth delay product of approximately 1 RTT. The window size used for WANs is 600000 Bytes.

- TCP delay ack timer is NOT set. Segments are acked as soon are they are received.

- Duration of simulation runs is 10 seconds for LANs and 20 seconds for WANs.

- All TCP sources start and stop at the same time. There is no processing delay, delay variation or randomization in any component of the simulation. This highlights the effects of TCP synchronization as discussed later.

### 3.2 Performance Metrics

The performance of TCP over UBR is measured by the efficiency and fairness which are defined as follows:

$$\text{Efficiency} = (\text{Sum of TCP throughputs})/(\text{Maximum possible TCP throughput})$$

The TCP throughputs are measured at the destination TCP layers. Throughput is defined as the total number of bytes delivered to the destination application divided by the total simulation time. The results are reported in Mbps.

The maximum possible TCP throughput is the throughput attainable by the TCP layer running over UBR on a 155.52 Mbps link. For 512 bytes of data (TCP maximum segment size), the ATM layer receives 512 bytes of data + 20 bytes of TCP header + 20 bytes of IP header + 8 bytes of LLC header + 8 bytes of AAL5 trailer. These are padded to produce 12 ATM cells. Thus, each TCP segment results in 636 bytes at the ATM Layer. From this, the maximum possible throughput = 512/636 = 80.5% = 125.2 Mbps approximately on a 155.52 Mbps link.

$$\text{Fairness Index} = (\Sigma x_i)^2 / (n \times \Sigma x_i^2)$$

Where $x_i$ = throughput of the $i$th TCP source, and $n$ is the number of TCP sources

The fairness index metric applies well to the n-source symmetrical configuration. For more general configurations with upstream bottlenecks, the max-min fairness criteria [5] can be used.

## 4 TCP over UBR

In its simplest form, an ATM switch implements a tail drop policy. When a cell arrives at the FIFO queue, if the queue is full, the cell is dropped, otherwise the cell is accepted. If a cell is dropped, the TCP source loses time waiting for the retransmission timeout. Even though TCP congestion mechanisms effectively recover from loss, the resulting throughput can be very low. It is also known that simple FIFO buffering with tail drop results in excessive wasted bandwidth. Simple tail drop of ATM cells results in the receipt of incomplete segments. When part of a segment is dropped at the switch, the incomplete segment is dropped at the destination during reassembly. This wasted bandwidth further reduces the effective TCP throughput.

We simulate 5 and 15 TCP sources with finite buffered switches. The simulations are performed with three values of switch buffer sizes both for LAN and WAN links. For WAN experiments, we choose buffer sizes of approximately $k$ times the bandwidth-delay product of the connection for $k$ = 1,2 and 3. Thus, we select WAN buffer sizes of 12000, 24000 and 36000 cells. These values are chosen because most feedback control mechanisms can achieve steady state in a fixed number of round trip times, and have similar buffer requirements for zero loss at the switch [9]. It is interesting to assess the performance of vanilla UBR in this situation. For LANs, 1 RTT × Bandwidth is a very small number (11 cells) and is not practical as the size for the buffer. For LAN links, the buffer sizes chosen are 1000, 2000, and 3000 cells. These numbers are closer to the buffer sizes of current LAN switches.

Column 4 of tables 2 and 3 show the efficiency and fairness values respectively for these experiments. Several observations can be made from these results.

Table 1: TCP over UBR: Buffer requirements for zero loss

| Number of Sources | Configuration | Efficiency | Fairness | Maximum Queue (Cells) |
|---|---|---|---|---|
| 5 | LAN | 1 | 1 | 7591 |
| 15 | LAN | 1 | 1 | 22831 |
| 5 | WAN | 1 | 1 | 59211 |
| 15 | WAN | 1 | 1 | 196203 |

- **TCP over vanilla UBR results in low fairness in both LAN and WAN configurtions.** This is due to TCP synchronization effects. TCP connections are synchronized when their sources timeout and retransmit at the same time. This occurs because packets from all sources are dropped forcing them to enter slow start phase. However, in this case, when the switch buffer is about to overflow, one or two connections get lucky and their entire windows are accepted while the segments from all other connections are dropped. All these connections wait for a timeout and stop sending data into the network. The connections that were not dropped send their next window and keep filling up the buffer. All other connections timeout and retransmit at the same time. This results in their segments being dropped again and the synchronization effect is seen. The sources that escape the synchronization get most of the bandwidth.

- **The default TCP maximum window size leads to low efficiency in LANs.** LAN simulations have very low effeciency values (less than 50%) while WAN simulations have higher effeciency values. For LANs, the the TCP receiver window size (65535 Bytes) corresponds to more than 1500 cells at the switch for each source. For 5 sources and a buffer size of 1000 cells, the sum of the window sizes is almost 8 times the buffer size. For WAN simulations, with 5 sources and a buffer size of 12000 cells, the sum of the window sizes is less than 6 times the buffer size. Moreover, the larger RTT in WANs allows more cells to be cleared out before the next window is seen. As a result, the WAN simulations have higher throughputs than LANs. For LAN experiments with smaller window sizes (less than the default), higher efficiency values are seen.

- **Efficiency typically increases with increasing buffer size.** Larger buffer sizes result in more cells being accepted before loss occurs, and therefore higher efficiency. This is a direct result of the dependence of the buffer requirements to the window sizes.

TCP performs best when there is zero loss. In this situation, TCP is able to fill the pipe and fully utilize the link bandwidth. During the exponential rise phase (slow start), TCP sources send out two segments for every segment that is acked. For N TCP sources, in the worst case, a switch can receive a whole window's worth of segments from N-1 sources while it is still clearing out segments from the window of the Nth source. As a result, the switch can have buffer occupancies of up to the sum of all the TCP maximum sender window sizes. This is especially true for connections with very small propagation delays. For large propagation delays, the switch has more time to clear out a segment before it sees the two segments which resulted from the ack.

Table 1 contains the simulation results for TCP running over UBR service with infinite buffering. The maximum queue length numbers give an indication of the buffer sizes required at the switch to achieve zero loss for TCP. The connections achieve 100% of the possible throughput and perfect fairness.

For the five source LAN configuration, the maximum queue length is 7591 cells = 7591 / 12 segments = 633 segments $\approx$ 323883 Bytes. This is approximately equal to the sum of the TCP window sizes (65535 $\times$ 5 Bytes). For the five source WAN configuration, the maximum queue length is 59211 cells = 2526336 Bytes. This is slightly less that the sum of the TCP window sizes (600000 × 5 = 3000000 Bytes). This is because the switch has 1 RTT to clear out almost 500000 bytes of TCP data (at 155.52 Mbps) before it receives the next window of data. In any case, the increase in buffer requirement is proportional to the number of sources in the simulation. The maximum queue is reached just when the TCP connections reach the maximum window. After that, the window stabilizes and TCP's self clocking congestion mechanism puts one segment into the network for each segment that leaves the network. **For a switch to guarantee zero loss for TCP over UBR, the amount of buffering required is equal to the sum of the TCP maximum window sizes for all the TCP connections.**

## 5 UBR+: Early Packet Discard

The Early Packet Discard (EPD) policy [1] has been suggested to remedy some of the problems with tail drop switches. EPD drops complete packets instead of partial packets. As a result, the link does not carry incomplete packets which would have been discarded during reassembly. A threshold R less than the buffer size, is set at the switches. When the switch queue length exceeds this threshold, all cells from any new packets are dropped. Packets which had been partly received before exceeding the threshold are still accepted if there is buffer space. In the worst case, the switch could have received one cell from all N connections before its buffer exceeded the threshold. To accept all the incomplete packets, there should be additional buffer capacity of the sum of the packet sizes of all the connections. Typically, the threshold R should be set to the buffer size − N × the maximum packet size, where N is the expected number of connections active at one time.

The EPD algorithm used in our simulations is the one suggested by [3, 10]. Column 5 of tables 2 and 3 show the efficiency and fairness respectively of TCP over UBR with EPD. The switch thresholds are selected so as to allow one entire packet from each connection to arrive after the threshold is exceeded. We use thresholds of Buffer Size − 200 cells in our simulations. 200 cells are enough to hold one packet each from all 15 TCP connections. This reflects the worst case scenario when all the fifteen connections have received the first cell of their packet and then the buffer occupancy exceeds the threshold.

Tables 2 and 3 show that **EPD improves the efficiency of TCP over UBR, but it does not improve fairness.** This is because EPD indiscriminately discards complete packets from all connections without taking into account their current rates or buffer utilizations. When the buffer occupancy exceeds the threshold, all new packets are dropped. The slight improvement in fairness in the LAN cases is because EPD can sometimes break TCP synchronization and in such cases only a few connections are dropped during congestion.

## 6 UBR+: Selective Drop using per-VC accounting

Per-VC accounting can be effectively used to achieve a greater degree of fairness among TCP connections. A VC that is using up excessive share of the throughput or buffer capacity can be penalized preferentially over another. The scheme presented here is a simpler version of the Fair Buffer Allocation scheme proposed in [6] and described in the next section. Selective Drop keeps track of the activity of each VC by counting the number of cells from each VC in the buffer. A VC is said to be active if it has at least one cell in the buffer. A fair allocation is calculated as the (current buffer occupancy) divided by (number of active VCs).

Let the buffer occupancy be denoted by $X$, and the number of active VCs be denoted by $N_a$. Then,

$$\text{Fair allocation} = X/N_a$$

The ratio of the number of cells of a VC in the buffer to the fair allocation gives a measure of how much the VC is overloading the buffer i.e., by what ratio it exceeds the fair allocation. Let $Y_i$ be the number of cells from $VC_i$ in the buffer. Then the Load Ratio of $VC_i$ is defined as

$$\text{Load Ratio of } VC_i = (\text{Number of Cells from } VC_i) / (\text{Fair allocation}) = Y_i \times N_a / X$$

If the load ratio of a VC is greater than a parameter Z, then new packets from that VC are dropped in preference to packets of a VC with load ratio less than Z. Thus, Z is used as a cutoff for the load ratio to indicate that the VC is overloading the switch.

Figure 3 shows the buffer management of the Selective drop scheme. For a given buffer size K (cells), the selective drop scheme assigns a static minimum threshold parameter R (cells). If the buffer occupancy X is less than or equal to this minimum threshold R, then no cells are dropped. If the buffer occupancy is greater than R, then the next new incoming packet of $VC_i$ is dropped if the load ratio of $VC_i$ is greater than Z.

We performed simulations to find the value of Z that optimizes the efficiency and fairness values. We first performed 5 source LAN simulations with 1000 cell buffers. We set R to 0.9 × the buffer size K. This ensured that there was enough buffer space accept incomplete packets during congestion. We experimented with values of Z = 2, 1, 0.9, 0.5 and 0.2. Z = 0.9 resulted in good results. A further simulation of Z around 0.9 shows that Z = 0.8 produces the best efficiency and fairness values for this configuration. For WAN simulations, any Z value between 0.8 and 1 produces good results. Tables 2,3 show the simulation results for the optimal performances of each scheme. The following observations can be made from the simulation results:

- **Selective Drop using per-VC accounting improves the fairness of TCP over UBR+EPD.** This is because cells from overloading connections are dropped in preference to underloading ones. As a result, Selective Drop is more effective in breaking TCP synchronization. When the buffer exceeds the threshold, only cells from overloading connections are dropped. This frees up some bandwidth and allows the underloading connections to increase their window and obtain more throughput.

- **Fairness and efficiency increase with increase in buffer size.**

- **Fairness decreases with increasing number of sources.**

# 7  UBR+: The Fair Buffer Allocation Scheme

The Fair Buffer Allocation Scheme proposed by [6] uses a smooth form of the parameter Z anc compares it with the Load ratio of a VC. To make the cutoff smooth, FBA uses the current load level in the switch. The scheme compares the load ratio of a VC to 1 + another threshold that determines how much the switch is congested. Let K be the buffer capacity of the switch in cells. For a given buffer size K, the FBA scheme assigns a static Minimum Threshold parameter R (cells). If the buffer occupancy X is less than or equal to this minimum threshold R, then no cells are dropped. When the buffer occupancy is greater than R, then upon the arrival of every new packet, the load ratio of the VC (to which the packet belongs) is compared to an allowable drop threshold calculated as Z×(1 + (K−X)/(X−R)). In this equation Z is a linear scaling factor. The next packet from $VC_i$ is dropped if

$$(X > R) \text{ AND } (Y_i \times N_a / X > Z((K - R)/(X - R)))$$

Figure 3 shows the switch buffer with buffer occupancies X relative to the minimum threshold R and the buffer size K where incoming TCP packets may be dropped.

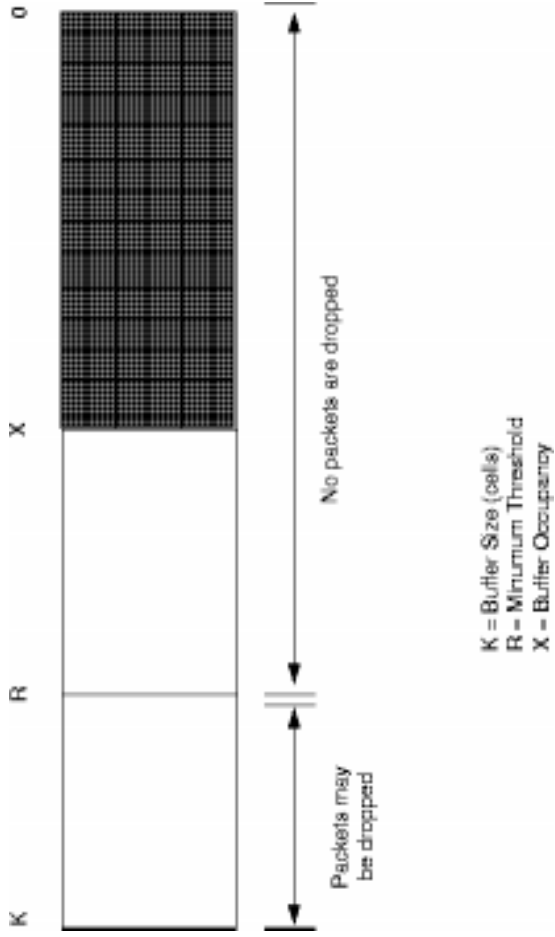

Figure 3: Selective Drop and FBA: Buffer Occupancy for drop

Note that when the current buffer occupancy X exceeds the minimum threshold R, it is not always the case that a new packet is dropped. The load ratio in the above equation determines if $VC_i$ is using more than a fair amount of buffer space. $X / N_a$ is used as a measure of a fair allocation for each VC, and $Z \times ((K - R)/(X - R))$ is a drop threshold for the buffer. If the current buffer occupancy ($Y_i$) is greater than this dynamic threshold times the fair allocation ($X / N_a$), then the new packet of that VC is dropped.

## 7.1 Effect of the minimum drop threshold R

The load ratio threshold for dropping a complete packet is $Z((K - R)/(X - R))$. As R increases for a fixed value of the buffer occupancy X, $X - R$ decreases, which means that the drop threshold $((K - R)/(X - R))$ increases and each connection is allowed to have more cells in the buffer. Higher values of R provide higher efficiency by allowing higher buffer utilization. Lower values of R should provide better fairness than higher values by dropping packets earlier.

## 7.2 Effect of the linear scale factor Z

The parameter Z scales the FBA drop threshold by a multiplicative factor. Z has a linear effect on the drop threshold, where lower values of Z lower the threshold and vice versa. Higher values of Z should increase the efficiency of the connections. However, if Z is very close to 1, then cells from a connection may not be dropped until the buffer overflows.

## 7.3 Effect of FBA parameters: Simulation results

We performed a full factorial experiment with the following parameter variations for both LANs and WANs. Each experiment was performed for N source configurations.

- Number of sources, N = 5 and 15.
- Buffer capacity, K = 1000 , 2000 and 3000 cells for LANs and 12000, 24000 and 36000 cells for WANs.

- Minimum drop threshold, R = 0.9*K , 0.5*K and 0.1*K.

- Linear scale factor, Z = 0.2 , 0.5 and 0.8.

A set of 54 experiments were conducted to determine the values of R and Z that maximized efficiency and fairness among the TCP sources. We sorted the results with respect to the efficiency and fairness values. The following observations can be made from the simulation results.

- **There is a tradeoff between efficiency and fairness.** The highest values of fairness (close to 1) have the lowest values of efficiency. The simulation data shows that these results are for low R and Z values. Higher values of the minimum threshold R combined with low Z values lead to slightly higher efficiency. Efficiency is high for high values of R and Z. Lower efficiency values have either R or Z low, and higher efficiency values have either of R or Z high. When R is low (0.1), the scheme can drop packets when the buffer occupancy exceeds a small fraction of the capacity. When Z is low, a small rise in the load ratio will result in its packets being dropped. This improves the fairness of the scheme, but decreases the efficiency especially if R is also low. **For configurations simulated, we foundthat the best value of R was about 0.9 and Z about 0.8.**

- **The fairness of the scheme is sensitive to parameters.** The simulation results showed that small changes in the values of R and Z can result in significant differences in the fairness results. With the increase of R and Z, efficiency shows an increasing trend. However there is considerable variation in the fairness numbers. We attribute this to TCP synchronization effects. Sometimes, a single TCP source can get lucky and its packets are accepted while all other connections are dropped. When the source finally exceeds its fair-share and should be dropped, the buffer is no longer above the threshold because all other sources have stopped sending packets and are waiting for timeout.

- **FBA improves both fairness and efficiency of TCP over UBR.** In general, the average efficiency and fairness values for FBA (for optimal parameter values) are higher than the previously discussed options. Tables 2,3 show the fairness and efficiency values for FBA switches with R = 0.9 and Z = 0.8.

# 8  UBR+: Summary

The previous sections have shown successive improvements for the UBR service category in ATM networks. We summarize the results in the form of a comparative analysis of the various options in UBR+. This summary is based on the choice of optimal parameters for the drop policies. For both selective drop and fair buffer allocation, the values of R and Z are chosen to be 0.9 and 0.8 respectively.

- **TCP achieves maximum possible throughput when no segments are lost.** To achieve zero loss for TCP over UBR, switches need buffers equal to the sum of the receiver windows of all the TCP connections.

- **With limited buffer sizes, TCP performs poorly over vanilla UBR switches.** TCP throughput is low, and there is unfairness among the connections. The coarse granularity TCP timer is an important reason for low TCP throughput.

- **UBR with EPD improves the throughput performance of TCP.** This is because partial packets are not being transmitted by the network and some bandwidth is saved. EPD does not have much effect on fairness because it does not drop segments selectively.

Table 2: UBR+: Comparative analysis (Efficiency)

| Config-uration | Number of Sources | Buffer Size (cells) | UBR | EPD | Selective Drop | FBA |
|---|---|---|---|---|---|---|
| LAN | 5 | 1000 | 0.21 | 0.49 | 0.75 | 0.88 |
| LAN | 5 | 2000 | 0.32 | 0.68 | 0.85 | 0.84 |
| LAN | 5 | 3000 | 0.47 | 0.72 | 0.90 | 0.92 |
| LAN | 15 | 1000 | 0.22 | 0.55 | 0.76 | 0.91 |
| LAN | 15 | 2000 | 0.49 | 0.81 | 0.82 | 0.85 |
| LAN | 15 | 3000 | 0.47 | 0.91 | 0.94 | 0.95 |
| WAN | 5 | 12000 | 0.86 | 0.90 | 0.90 | 0.95 |
| WAN | 5 | 24000 | 0.90 | 0.91 | 0.92 | 0.92 |
| WAN | 5 | 36000 | 0.91 | 0.81 | 0.81 | 0.81 |
| WAN | 15 | 12000 | 0.96 | 0.92 | 0.94 | 0.95 |
| WAN | 15 | 24000 | 0.94 | 0.91 | 0.94 | 0.96 |
| WAN | 15 | 36000 | 0.92 | 0.96 | 0.96 | 0.95 |

Table 3: UBR+: Comparative analysis (Fairness)

| Config-uration | Number of Sources | Buffer Size (cells) | UBR | EPD | Selective Drop | FBA |
|---|---|---|---|---|---|---|
| LAN | 5 | 1000 | 0.68 | 0.57 | 0.99 | 0.98 |
| LAN | 5 | 2000 | 0.90 | 0.98 | 0.96 | 98 |
| LAN | 5 | 3000 | 0.97 | 0.84 | 0.99 | 0.97 |
| LAN | 15 | 1000 | 0.31 | 0.56 | 0.76 | 0.97 |
| LAN | 15 | 2000 | 0.59 | 0.87 | 0.98 | 0.96 |
| LAN | 15 | 3000 | 0.80 | 0.78 | 0.94 | 0.93 |
| WAN | 5 | 12000 | 0.75 | 0.94 | 0.95 | 0.94 |
| WAN | 5 | 24000 | 0.83 | 0.99 | 0.99 | 1 |
| WAN | 5 | 36000 | 0.86 | 1 | 1 | 1 |
| WAN | 15 | 12000 | 0.67 | 0.93 | 0.91 | 0.97 |
| WAN | 15 | 24000 | 0.82 | 0.92 | 0.97 | 0.98 |
| WAN | 15 | 36000 | 0.77 | 0.91 | 0.89 | 0.97 |

- **UBR with selective packet drop using per-VC accounting improves fairness over UBR+EPD.** Connections with higher buffer occupancies are more likely to be dropped in this scheme. The efficiency values are similar to the ones with EPD.

- **UBR with the Fair Buffer Allocation scheme can improve TCP throughput and fairness.** There is a tradeoff between efficiency and fairness and the scheme is sensitive to parameters. We found R = 0.9 and Z = 0.8 to produce best results for our configurations.

- **TCP synchronization is an important factor that effects TCP throughput and fairness.** Vanilla UBR and EPD are ineffective in breaking TCP synchronization because they drop packets from all connections. Selective feedback schemes are needed to break synchronization effects. Some values of FBA parameters are successful in breaking TCP synchronization, and for these values, we see high values of efficiency and fairness. Some other papers on TCP over UBR have broken TCP synchronization by artificially staggering the TCP sources or introducing some randomness in the simulation. This situation may not reflect TCP sources in the real world and we have chosen to not introduce any artificial randomness to break synchronization.